\documentclass[aps,prl,twocolumn,groupedaddress]{revtex4}
\usepackage{graphicx}
\begin{document}

\title{Quantum phase transition in the Hartree-Fock wave function of the hydrogen molecule}

\author{Mikhail V. Ivanov}
\email{mivanov@mi1596.spb.edu}
\affiliation{Institute of Precambrian Geology 
and Geochronology, Russian Academy of Sciences, Nab. Makarova 2, 
St. Petersburg 199034, Russia }

\date{\today}

\begin{abstract}
Precise solutions of the Hartree-Fock equations for the ground state of the hydrogen molecule are 
obtained for a wide range of internuclear distances $R$ by means of a two-dimensional fully numerical mesh 
computational method. The spatial parts of the single-electron wave functions are found to be coinciding 
for $R < 2.30$~a.u. At larger distances they become different and as $R\rightarrow\infty$ each of them takes the form 
corresponding to a separate atom. This quantum phase transition occurring at $R = 2.30$~a.u. gives a natural 
boundary between a delocalized behavior of electrons in the molecule and their distribution over separate 
atoms. This phenomenon can be also considered as an analog of the Wigner crystallization or the Mott 
transition on the level of a single molecule.
\end{abstract}


\maketitle

In recent years multiple studies of quantum phase transitions 
were carried out. These studies are significant for 
understanding of many complicated phenomena in solid 
states, clusters and so on. As an example, studies of the 
Wigner crystallization in two-dimensional systems can be 
presented \cite{2d}. Many of these effects are associated with 
symmetry breakings in the Hartree-Fock ground states. On 
the other hand, spontaneous symmetry breakings in 
molecular systems were reported mainly in a negative 
context \cite{molasymm}. 
In this communication we study the simplest 
molecule with more than one electron, 
i.e. the ${\rm H_2}$ molecule, 
for arbitrary internuclear distances $R$. This molecule was 
investigated in a number of works by means of many precise 
methods (see \cite{wolniewicz}). 
On the other hand, there are no studies of 
precise solutions of the Hartree-Fock equations 
for the ${\rm H_2}$ 
molecule for arbitrary internuclear distances. We present 
such calculations and show that at some $R = R_{\rm cr}$ the wave 
function of the ground state undergo a quantum phase 
transition associated with a spontaneous symmetry breaking. 
This transition separates a phase with two fully delocalized 
electrons with equal spatial parts of their wave functions and 
a phase with two electrons localized on separate atoms 
($R > R_{\rm cr}$). 
This phenomenon is very similar to the forming of 
Wigner crystals or Mott transitions in the solid state context. 
Due to this phase transition the solution 
of the Hartree-Fock equations has a reasonable 
physical meaning for arbitrary $R$ values, that does not take 
place in a traditional approach with single-electron wave 
functions having the symmetry of the molecule as the whole 
at all the $R$ values \cite{wilson}. 
This result allows us to expect, that, as well 
as in the solid state physics, some symmetry breakings in 
more complicated molecules and similar systems also could 
be not computational artifacts \cite{molasymm} 
but have an appropriate 
physical meaning.

We solve a system of the Hartree-Fock (HF) equations by a 
fully numerical two-dimensional finite-difference (mesh) 
method \cite{iva,ivaschm,ivabe}. The equations are presented in cylindrical 
coordinates $(\rho,\phi,z)$ with the axis $z$ coinciding with the 
molecular axis. The point $z = 0$ corresponds to the center of 
gravity of the molecule. For each single-electron wave 
function a definite value of the magnetic quantum number is 
supposed ($m = 0$ for the ground state, considered below), so 
that a numerical solution is carried out on the plane $(\rho,z)$. 
Thus, the wave function of the spin singlet ground state has 
the form 
\begin{eqnarray}
\Psi = \psi_1(\rho_1,z_1)\psi_2(\rho_2,z_2)\nonumber
\end{eqnarray}
where $\psi_1$ and $\psi_2$ are solutions 
of the Hartree-Fock equations 
for the first and second electrons. The corresponding energy 
of the electron system we denote as $E_{\rm e}$. 
When $\psi_1\neq\psi_2$ this 
wave function is not an eigenfunction of the spin operator 
${\rm \bf \hat S}^2$. 
In this case it is possible to consider also 
a spin-symmetrised two-determinant wave function
\begin{eqnarray}
\Psi^{\rm symm}=
[\psi_1(\rho_1,z_1)\psi_2(\rho_2,z_2)+
\psi_2(\rho_1,z_1)\psi_1(\rho_2,z_2)]\nonumber
\end{eqnarray}
and the corresponding energy $E{\rm_e^{symm}}$. The latter wave 
function is an eigenfunction of the operator 
${\rm \bf \hat S}^2$  as $\psi_1\neq\psi_2$. 

A traditional approach to solution of the molecular 
Hartree-Fock problem employs fully delocalized molecular orbitals. 
In our case this means $\psi_1 = \psi_2$. 
This allows obtaining reasonable results only for relatively small 
$R$ values. 
For large distances atomic orbitals 
strictly localized near corresponding nuclei are sometimes considered. 
Of course, every combination of such approaches 
gives rise to a problem of intermediate 
values of the parameter $R$. In our calculations we are free from 
limitations associated with a choice of basis functions and do 
not require this or that way of behavior from $\psi_1$ and $\psi_2$.
They are direct solutions of the initial equations. 

\begin{figure}
\includegraphics[width=8.5cm,clip]{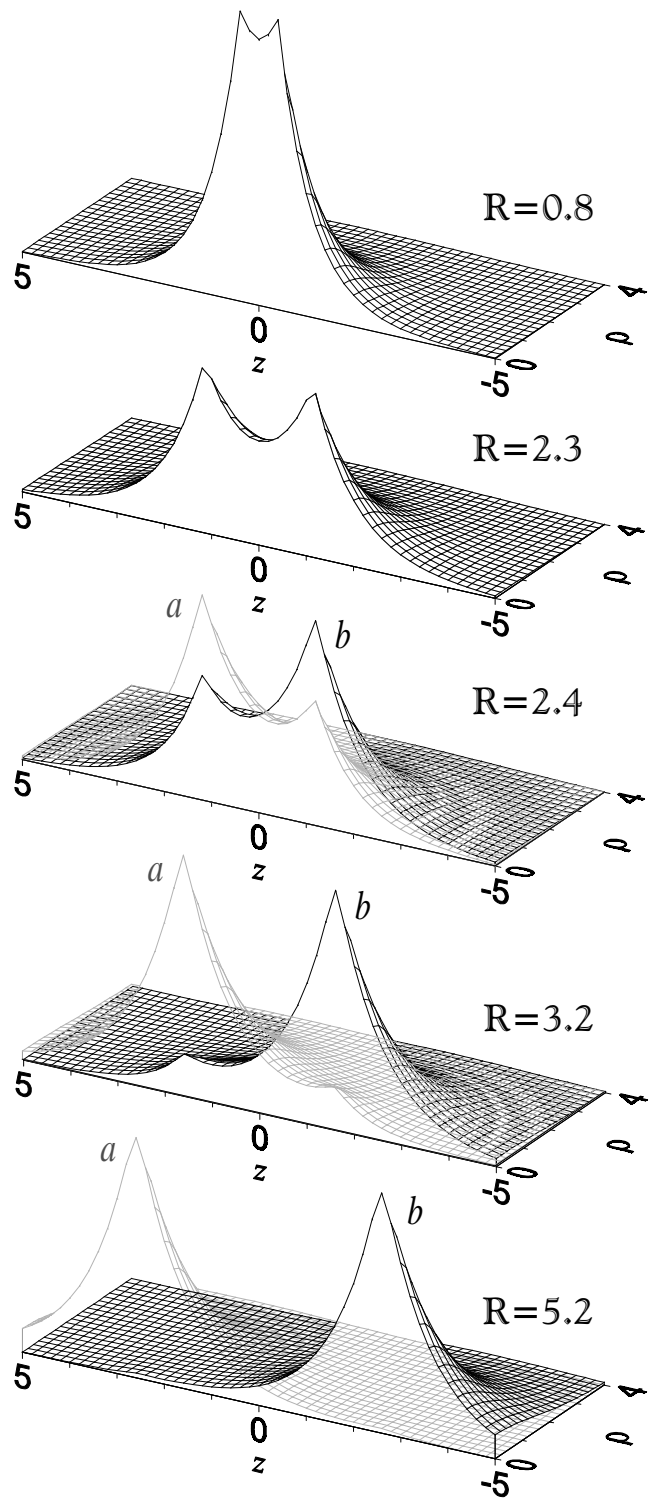}%%
\caption{Single electron Hartree-Fock wave functions of the 
hydrogen molecule for different internuclear distances $R$. For 
$R > R_{\rm cr}\approx 2.3$~a.u. $a$ (light-gray lines) and $b$ (black lines) are 
the functions with electron densities concentrated at the left 
and right nuclei respectively. For $R < R_{\rm cr}$ these wave 
functions coincide.
\label{figur1}}
\end{figure}

\begin{figure}
\includegraphics[width=8.5cm,clip]{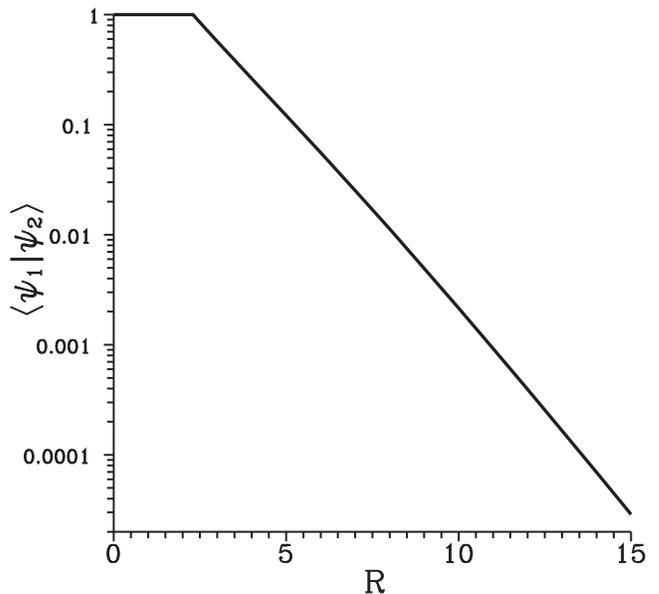}%%
\caption{Overlap integral for the spatial parts of the 
single-electron wave functions of the HF hydrogen molecule 
as a function of the internuclear distance (a.u.).
\label{figur2}}
\end{figure}

\begin{figure}
\includegraphics[width=8.5cm,clip]{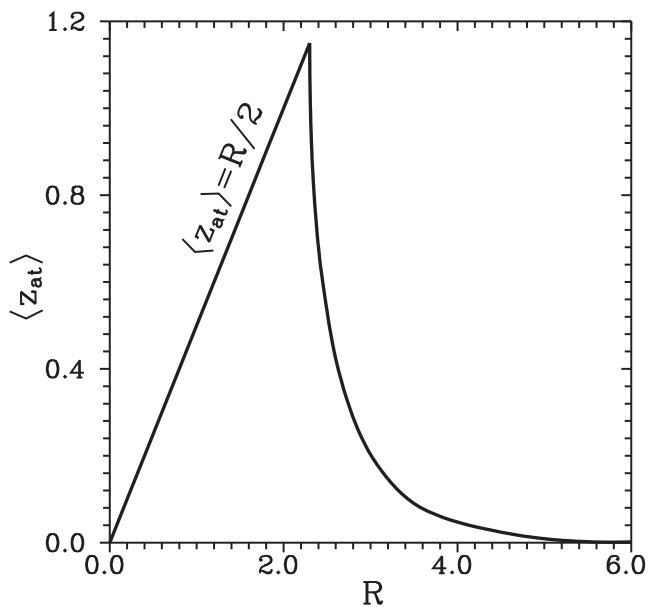}%%
\caption{Dipole moment of a single electron wave function 
(a.u.) related to the position of the corresponding nucleus. 
The dependence on the internuclear distance.
\label{figur3}}
\end{figure}

The results obtained in our numerical solution of the 
Hartree-Fock equations are presented in the Table~\ref{tab:tab1} 
and Figures~\ref{figur1}-\ref{figur3}. All the 
data are given in atomic units. As one can see, the solutions 
have very different characters for $R < 2.3$ and for $R > 2.3$. 
Several plots of the spatial parts of the single-electron wave 
functions $\psi_1$ and $\psi_2$ for different $R$ are given in 
Figure~\ref{figur1}. For $R > 2.3$ we have $\psi_1\neq\psi_2$ 
(functions $a$ and $b$ in the plots), 
and with growing $R$ these solutions transform into wave 
functions of separate hydrogen atoms. On the other hand, at 
$R < 2.3$ the functions $\psi_1$ and $\psi_2$ fully coincide. 
This relation 
between $\psi_1$ and $\psi_2$ as well as the existence of the critical 
point $R = R_{\rm cr}$ for solutions of the Hartree-Fock equations can 
be illustrated by the behavior of their overlap integral 
$\left<\psi_1|\psi_2\right>$ 
given in Table~\ref{tab:tab1} and Figure~\ref{figur2}. 
For $R < R_{\rm cr}$ an exact equality 
$\left<\psi_1|\psi_2\right> = 1$ takes place, whereas for $R$ above the critical 
point the function $\left<\psi_1|\psi_2\right>(R)$ shows a near to an exponential 
decrease with increasing $R$ values. Our calculations in a 
vicinity of the critical point allow to estimate its position as 
$R_{\rm cr} = 2.30$ or, more precisely, $R_{\rm cr} = 2.29999$. 
For $R < R_{\rm cr}$ we have 
$\psi_i(\rho,z) = \psi_i(\rho,-z)$ ($i = 1,2$), 
whereas for $R > R_{\rm cr}$ this symmetry 
condition does not take place and only condition 
$\psi_1(\rho,z) = \psi_2(\rho,-z)$ is preserved. It should be emphasized, that 
the transition from the single-electron wave functions, 
symmetric with respect to the center of the molecule, to the 
non-symmetric ones is an intrinsic property of the 
Hartree-Fock equations for the hydrogen molecule. 
Thus, for the hydrogen molecule at 
$R > R_{\rm cr}$ a spontaneously broken symmetry in the 
Hartree-Fock wave function takes place. A similar effect has been 
found previously for the beryllium atom~\cite{ivabe}. 
From the point 
of view of the numerical techniques the Hartree-Fock 
iterations can be started from initial wave functions with 
arbitrary small deviations from exact $z = 0$ parity. The final 
result of iterations does not depend on the degree of this 
initial asymmetry (this could be simply round-off errors) and 
has the form presented in Figure~\ref{figur1}. 

\begin{table}
\caption{Physical parameters of the Hartree-Fock hydrogen 
molecule dependent on the internuclear distances. Atomic 
units.}
\begin{ruledtabular}
\begin{tabular}{lllllll}
$R$&$E{\rm_e}$&$E{\rm_e^{symm}}$&$\left<\psi_1|\psi_2\right>$\\ \noalign{\hrule} 
$0.0$&$-2.861679998 $&&$1$\\
$0.2$&$-2.7608175   $&&$1$\\
$0.4$&$-2.5798578   $&&$1$\\
$0.6$&$-2.396660751 $&&$1$\\
$0.8$&$-2.23074400  $&&$1$\\
$1.0$&$-2.085138396 $&&$1$\\
$1.2$&$-1.958359581 $&&$1$\\
$1.3$&$-1.901255170 $&&$1$\\
$1.4$&$-1.847915286 $&&$1$\\
$1.5$&$-1.798039362 $&&$1$\\
$1.6$&$-1.751347985 $&&$1$\\
$1.8$&$-1.666513946 $&&$1$\\
$2.0$&$-1.591619847 $&&$1$\\
$2.2$&$-1.525186768 $&&$1$\\
$2.3$&$-1.494754132 $&$-1.494754253$&$0.9999997   $\\
$2.4$&$-1.466869931 $&$-1.47724495 $&$0.92138966  $\\
$2.6$&$-1.41976886  $&$-1.44212958 $&$0.784583061 $\\
$2.8$&$-1.381746342 $&$-1.408258544$&$0.66916303  $\\
$3.0$&$-1.35055131  $&$-1.37663659 $&$0.57149223  $\\
$3.5$&$-1.29277168  $&$-1.31056772 $&$0.38687543  $\\
$4.0$&$-1.252894522 $&$-1.262622240$&$0.262823069 $\\
$5.0$&$-1.200485964 $&$-1.202743366$&$0.121438978 $\\
$6.0$&$-1.166747578 $&$-1.167204293$&$0.055683757 $\\
$7.0$&$-1.14287045  $&$-1.14295695 $&$0.025207015 $\\
$8.0$&$-1.12500216  $&$-1.12501783 $&$0.011249534 $\\
$9.0$&$-1.11111146  $&$-1.11111420 $&$0.004951520 $\\
$10.$&$-1.10000006  $&$-1.10000053 $&$0.00215182  $\\
$11.$&$-1.09090911  $&$-1.09090918 $&$0.0009244618$\\
$12.$&$-1.08333334  $&$-1.08333336 $&$0.0003931164$\\
$13.$&$-1.07692309  $&$-1.07692309 $&$0.0001656485$\\
$14.$&$-1.0714286   $&$-1.0714286  $&$0.0000692334$\\
$15.$&$-1.0666667   $&$-1.0666667  $&$0.0000287261$\\
\end{tabular}
\end{ruledtabular}
\label{tab:tab1}
\end{table}

The Hartree-Fock electron energies of the hydrogen 
molecule as well as numerical values of the $\left<\psi_1|\psi_2\right>$ are 
presented in Table~\ref{tab:tab1}. 
For $R < R_{\rm cr}$ the Hartree-Fock electron 
energy is an energy calculated on a one-determinant wave 
function of the whole system $E{\rm_e}$. 
Available precise HF result 
for the near to the equilibrium distance $R = 1.4$ 
($E{\rm_e} = -1.8479152858$, \cite{kobus}) coincide with our one. 
Our energy for 
$R = 0$ can be considered as more precise than the well known 
result for the helium atom $E{\rm_e} = -2.8616799$ \cite{clementy}. 
For $R > R_{\rm cr}$ it 
is possible to calculate the spin-symmetrised energy 
$E{\rm_e^{symm}}$ 
also presented in the table. 
This energy lies lower than $E{\rm_e}$ 
except the cases $\left<\psi_1|\psi_2\right> = 0$ 
and $\left<\psi_1|\psi_2\right> = 1$, when both 
energies evidently coincide. 
 
As a discussion it is expedient to note that the result obtained 
above contradicts to a frequently encountered opinion, that 
even the simplest of molecules, the hydrogen molecule, 
dissociates incorrectly within the Hartree-Fock 
approximation \cite{wilson}. 
When we do not impose some artificial 
conditions (i.e. the symmetry of the molecule as a whole) 
on single-electron wave functions, the consistent 
Hartree-Fock approach allows obtaining quite reasonable 
solutions for arbitrary internuclear separations. An exception 
are Van der Waals forces, which cannot be obtained in the 
Hartree-Fock approximation. In the limit $R\rightarrow\infty$ 
our solution 
describes two separate atoms (H + H) with electronic wave 
functions $\psi_1$ and $\psi_2$ respectively. 
On the other hand, when 
establishing symmetry conditions 
$\psi_i(\rho,z) = \psi_i(\rho,-z)$ or/and 
$\psi_1 = \psi_2$, we have in the limit $R\rightarrow\infty$ 
a traditional non-physical Hartree-Fock result (see \cite{wilson}), 
which does not correspond 
to the ground states of the systems H + H or ${\rm H^-} + p$.

Some different interpretation of our result could be obtained 
when we consider two separate hydrogen atoms with 
anti-parallel spins of electrons and trace their wave functions 
when shortening the distance between them. Let us examine, 
for example, a dipole moment $\left<z_{\rm at}\right>$ of a hydrogen atom, when 
it nears to another hydrogen atom. This dependence is 
presented in Figure~\ref{figur3}. 
As well as the overlap integral this 
value increases nearly exponential with reducing $R$ at large 
distances, but its behavior near the critical point is even more 
pronounced because its derivative becomes infinite here. For 
$R < R_{\rm cr}$ an evident for delocalized electrons relation 
$z_{\rm at} = R/2$ takes place.

It is natural to ask a question about behavior of 
Hartree-Fock wave functions of other states of the hydrogen 
molecule at large $R$. A detailed investigation lies outside the 
scope of this communication and we outline here the main 
characteristics of states, which could be considered from 
some points of view as resembling the ground state 
configurations of H + H or ${\rm H^-} + p$ 
in the limit $R\rightarrow\infty$. In the 
notation of a united atom they are configurations 
$1s\uparrow2p_0\downarrow$, 
$1s\uparrow2p_0\uparrow$ and 
$2p_0^2$, along with the ground state $1s^2$ considered above. 
There is no quantum phase transitions in both singlet and 
triplet configurations $1s2p_0$. 
As $R\rightarrow\infty$ the energy of the 
single-determinant wave function for the $1s\uparrow2p_0\downarrow$ (as well 
as for the $2p_0^2$ when 
$\psi_i(\rho,z) = -\psi_i(\rho,-z)$ and $\psi_1 = \psi_2$) has a 
non-physical limit coinciding with that for $1s^2$ under 
condition $\psi_1 = \psi_2$. On the other hand, due to exchange terms 
the energy of the triplet $1s2p_0$ configuration has the limit 
corresponding to the ground state of two separate hydrogen 
atoms with parallel electron spins. A spin-symmetrized 
two-determinant wave function of the singlet $1s2p_0$ tends to that 
of the system ${\rm H^-} + p$ as $R\rightarrow\infty$. 
The most complicated is the 
behavior of the configuration $2p_0^2$, when we do not impose 
additional symmetries for it. It retains its initial atomic 
symmetries up to $R \approx 2.97$. 
For larger $R$ values the wave functions 
lose their parity with respect to the plane $z = 0$, but they 
remain equal up to $R \approx 5.9$. 
At this point the second phase 
transition, leading to $\psi_1\neq\psi_2$ takes place. 
However, this series of transformations does not lead to a 
reasonable physical result as $R\rightarrow\infty$.

The effect presented above can be considered in a variety of 
aspects, and analogies with several circles of phenomena 
could be established. One of these items is a problem of 
localization of electrons in molecules (discussions and 
references see in ref.~\cite{wilson}, some more recent approaches see 
e.g. in ref.~\cite{dobrodey}). Our result gives a natural boundary, 
separating two regimes: 1. Delocalized molecular orbitals and 
2. Electrons localized on separate atoms (where, e.g. the 
value $1 - \left<\psi_1|\psi_2\right>$ 
could be employed as a proper characteristic of 
the localization). Thus, we can consider our result as a 
solution of this problem for the hydrogen molecule, and a 
possibility of similar solutions for more complicated 
molecules can be expected.

Another evident analogy are the Mott transition, Wigner 
localization and related topics like the Hubbard model. It is 
possible to consider the effect presented above as an analog 
of the Wigner localization or Mott transition on the level of a 
single molecule. Following steps in this direction have to be, 
of course, investigations of systems with more than two 
electrons, correlated systems and so on. Our calculations are 
carried out in the Hartree-Fock approximation which can be 
considered as obsolete and not precise for the hydrogen 
molecule. On the other hand, precise solutions of 
Schr\"odinger equations for more or less complex systems are 
unavailable and the methods of the Hartree-Fock level are 
the best of existing tools for them. Our result shows that for a 
simplest non-ionized molecule the symmetry breaking in a 
solution of the HF equations allows obtaining physically 
correct results for arbitrary internuclear distances and could 
give a simple insight into various more complicated 
problems.

{}

\end{document}